\documentclass[epsf,runningheads]{cl2emult}

\usepackage{makeidx}  
\usepackage{graphicx} 
\usepackage{subeqnar} 
\usepackage{multicol} 
\usepackage{cropmark} 
\usepackage{lnp}      

\begin{document}
\title*{ISOGAL survey of Baade's Windows}
\titlerunning{ISOGAL survey of Baade's Windows}
\author{Ian S. Glass\inst{1}
\and D.R. Alves\inst{2}
\and the ISOGAL and MACHO teams}
\authorrunning{I.S. Glass, D.R. Alves et al}
%
%
\institute{S.A. Astronomical Observatory,
           PO Box 9,
           Observatory 7935,
           South Africa  
\and Space Telescope Science Institute,
           Baltimore,
           MD 21218,
           USA}

\maketitle              

\begin{abstract}
The Baade's Windows of low obscuration towards the inner parts of the
Galactic bulge represent ideal places in which to develop an understanding
of the ISOGAL colour-magnitude diagrams. Unlike the case for the solar 
neighbourhood, their contents are at a uniform distance from the Sun, 
affected only by the finite thickness of the Bulge. 

The objects detected in the ISOGAL survey are found to be late-type M-giants 
at the red giant tip or on the Asymptotic Giant Branch (AGB). The ISOGAL 
colour-magnitude diagrams show that mass-loss starts at about M4 
and increases towards later types. Many non-Miras have mass-loss rates 
similar to shorter-period Miras. 

The visible counterparts of the ISOGAL sources have been identified in the 
database of the MACHO gravitational lensing survey. A first report of this 
work is included here. It is found that nearly all the ISOGAL sources are 
semi-regular variables (SRVs), which are many times more numerous than 
Miras. Their stellar luminosities increase with period. Based on a simple
interpretation of the photometry, mass-loss rates from about 10$^{-9}$ 
$M_{\odot}$ yr$^{-1}$ to 10$^{-7}$ $M_{\odot}$ yr$^{-1}$ are found for SRVs 
with periods in excess of $\sim$60 days.

\end{abstract}

\section{Introduction}

The ISOGAL survey (see Omont, this volume) has covered a large number of 
heavily obscured fields in the inner galaxy in the mid-infrared, using the 
ISOCAM filters LW2 (5.5--8\,$\mu$m) and LW3 (12--18\,$\mu$m). Very little
information is available concerning individual stars in most of the survey
areas and it was decided to include the relatively well-studied Baade's 
Windows (BW) of low obscuration, namely NGC\,6522 ($l$=+1$^{\circ}$, 
$b$=--3.8$^{\circ}$) and Sgr\,I ($l$=1.37$^{\circ}$, $b$=--2.63$^{\circ}$), 
whose $A_V$ is about 1.5, for comparison purposes. The M-type stellar 
content of the NGC\,6522 field has been surveyed by Blanco, McCarthy and 
Blanco (1984) and Blanco (1986). Frogel \& Whitford (1987) have presented 
near-IR photometry of many stars. The census of Mira ($V$ amplitude $>$ 2.5 
mag) variables in these fields is complete and their periods have been 
found by Lloyd Evans (1976, photographic infrared) and Glass et al.\ 
(1995, {\it JHKL} region). The $K$ and bolometric magnitudes of the Miras 
obey period-luminosity relations. No carbon-type AGB stars have been 
found in these fields. 

The results of the ISOGAL BW survey have recently been presented by Glass, 
Ganesh et al.\ (1999). In the two ISOGAL fields, each of 15 $\times$ 15 
arcmin$^2$, a total of 1,193 objects were found. The survey is believed 
to be complete to a level of 5\,mJy in both bands, corresponding to 
[7] = 10.64 and [15] = 8.99 mag. The photometric errors are $<$ 0.2 mag 
for bright sources, rising to $<$ 0.4 mag for faint ones. The sensitivity 
and spatial resolution of ISOCAM are about two orders of magnitude better 
than with IRAS. At the faint end, the density of sources approaches the 
confusion limit.

The principal conclusions of the survey are illustrated by Fig.\ 1. There is 
a continuous sequence of increasing [15] mag with [7] -- [15] 
colour. Much of the scatter is due to the distribution in 
depth of the Bulge, as found for the Mira log\, $P$ -- $K$ 
relation ($\sigma$ = 0.35; see Glass et al., 1995). Making use of the 
known spectral types of many objects, the data are interpreted as 
evidence for increasing dust emission with increase of stellar 
luminosity and decrease of temperature. Substantial numbers of 
stars in these fields have luminosities and mass-loss rates similar 
to those of the shorter-period Miras.

\begin{figure}
\centering
\includegraphics[width=.6\textwidth]{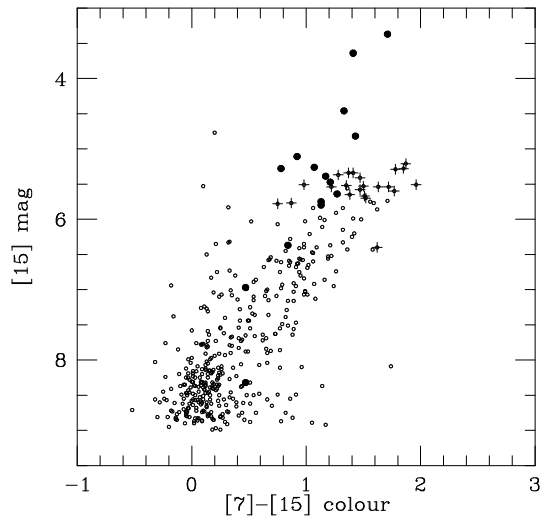}
\caption[]{ISOGAL colour-magnitude diagram for sources detected in the
NGC\,6522 and Sgr\,I Baade's Windows. A sequence of
increasing 15\,$\mu$m flux (flux arising in part from dust emission) with 
[7]--[15] colour (i.e., dust relative to photosphere), starts at 
the tip of the RGB in the bottom left corner and ends with the Miras at top 
right (solid points). Many other stars in the diagram, such as those 
with crosses, have luminosities and mass-loss rates similar to the 
shorter-period Miras. There is some contamination from foreground stars with
small [7] - [15] colour.}
\label{eps1}
\end{figure}

The ISOGAL 7\,$\mu$m flux of a late-type star with an optically thin dust 
shell arises from its photosphere, while the 15\,$\mu$m flux arises from
a combination of the photosphere and the dust. This is illustrated in 
Fig.\ 2, an opacity-sampled stellar atmosphere calculation including 
``astronomical silicate" dust, taken from Aringer et al.\ (1999). The 
bandpasses of the ISOCAM LW2 and LW3 filters are superimposed. It will be 
noticed that the 7\,$\mu$m band is hardly affected by dust emission, but
exhibits absorption features, probably due to water vapour and SiO. 
The 15$\mu$m band, on the other hand, may be dominated by silicate 
emission, which can greatly exceed that from the photosphere in this region.

\begin{figure}
\centering
\includegraphics[width=0.6\textwidth]{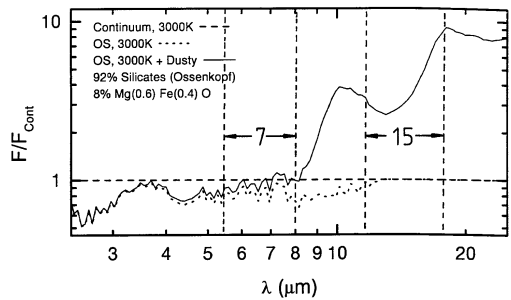}
\caption[]{Model of a 3000K stellar atmosphere with an ``astronomical
silicate" dust shell, taken from Aringer et al (1999), with ISOCAM filter
limits marked. The fluxes have been normalized to a 3000\,K blackbody. 
Note that the 7$\mu$m ISOGAL filter mainly measures a photospheric flux, 
while the 15$\mu$m filter is mainly sensitive to dust.}
\label{fig2}
\end{figure}

\section{Correlation with MACHO}

The Baade's Windows fields form part of the Bulge area that was surveyed 
nightly for six seasons of $\sim$ 250 days by the MACHO gravitational 
lensing project. The MACHO observations were made in two bands, $v$ and $r$, 
at effective wavelengths around 500\,nm and 700\,nm, and were transformed 
using formulae from Alcock et al.\ (1999) to Kron-Cousins $V, R$. 
The completeness and sensitivity to small-amplitude variations of MACHO 
and similar CCD-based surveys is much greater than in all previous work
which depended on photographic techniques. In particular, large numbers 
of small-amplitude SRVs in the solar neighbourhood might be found if data 
of MACHO quality were available.
 
Counterparts of the ISOGAL sources were sought within a radius of 3 
arcsec of their nominal positions. Since it is known that the ISOGAL 
sources are red giants, or else very bright early-type stars, only 
MACHO stars with $V$ $<$ 13.5 + 4.67$(V-R)$ were considered as candidates, 
with $(V-R)$ taken to be 0.5 in cases where only one colour was available. 
This left 40,000 objects in the two fields to select from. There were 
904 positional matches. The distributions of distance residuals for the 
two fields, together with, as a test, random matches produced when the 
MACHO star coordinates were displaced by 15 arcsec, show that spurious
matches should not exceed about 10\% in NGC\,6522 and about 20\% in Sgr\,I,
which is a denser field. Sources which were not matched fell on gaps in 
the MACHO detector mosaic or, in a few cases, were too bright to be included
in the MACHO database. A total of 332 stars had photometry at $V$, $R$, 
7\,$\mu$m and 15\,$\mu$m and these were analysed further.

\section{Periods and light curves}

Almost all the 332 selected counterparts show variability at some level, and
may be classified as semi-regular variables. Twenty-eight stars, most 
of them foreground bright objects, were rejected for saturation effects in
their MACHO lightcurves (these included two Miras). Five of the $\sim$14 
known Miras in the ISOGAL fields (see Glass, Ganesh et al., 1999) were 
recovered; the remaining 7 had already been rejected at the matching stage.
Fourier amplitude spectra were calculated for all 332 members of the 
sample. In general, a short period (15 to 200 days) could be identified, 
but, for many sources, slower variability, not necessarily periodic, 
was also evident. Because of the seasonality of the data, each season 
was also analysed separately and the Fourier amplitude spectra were 
summed before searching for the most significant periods. Work is 
continuing on the period-finding, which may still be subject to revision.

\begin{figure}
\centering
\includegraphics[width=1.0\textwidth]{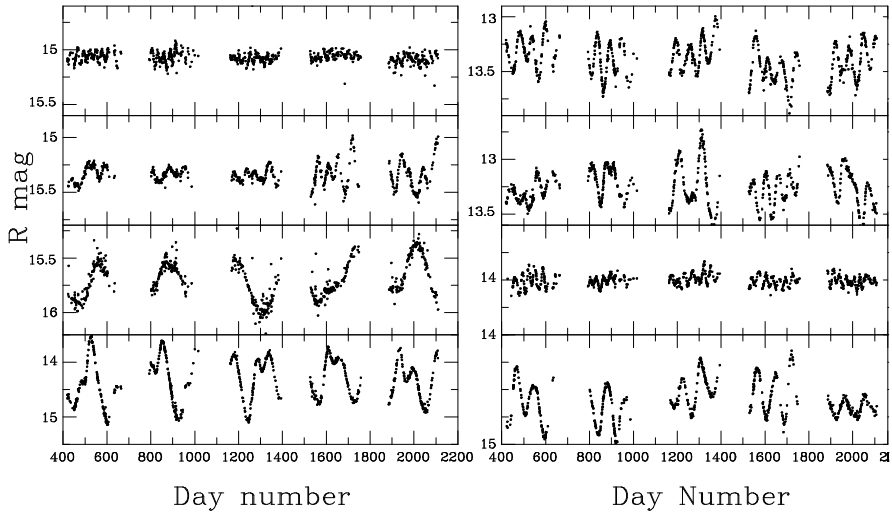}
\caption[]{Sample MACHO light curves of SRVs.}
\label{fig3}
\end{figure}

\begin{figure}
\centering
\includegraphics[width=0.6\textwidth]{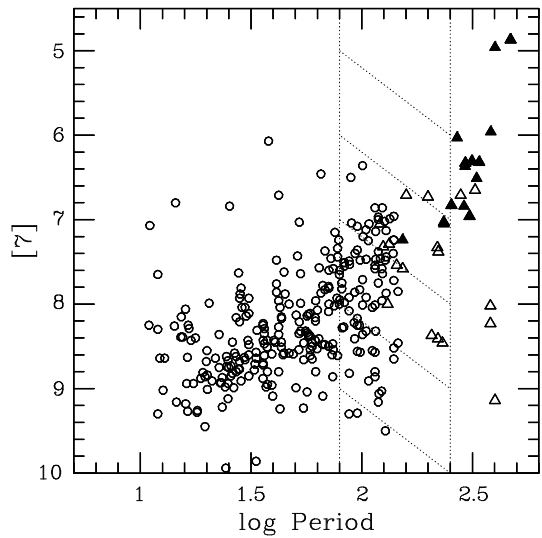}
\caption[]{Log $P$, [7] plot for cross-identified objects. The open points
are the main periods identified. Open triangles are separately identified
long periods and solid triangles are Mira (large-amplitude) variables, some
of which were saturated in the MACHO data. Periods for these cases were
taken from Glass et al.\ (1995). The periods in the hatched area may be subject 
to some revision.}
\label{fig4}
\end{figure}

The light curves (see Fig.\ 3) and periods are similar to those found 
by Wood et al (1999) for AGB stars in the Large Magellanic Cloud. 
As in the LMC, the SRVs outnumber the Miras by a large factor 
($\sim$ 20 in this case).  However, although most SRV light curves 
show clear evidence for variability in the 15--200 day range, as stated, 
only a few also show the longer periods around 300--400 days that seem 
to be common in the LMC (Type D in Fig.\ 2 of Wood et al.\, 1999).

In the log $P$, [7] diagram (Fig.\ 4), a clear period-luminosity correlation 
is seen for the SRVs (periods below log\,$P$ = 2.2) with a steeper slope 
for the Miras. The stars with periods longer than 200 days include the
Miras, which do not show simultaneous shorter periods. One star 
shows a single period of around 300 days and a 7$\mu$m luminosity 
appropriate to a Mira, but is not a Mira. The group of three stars around 
log $P$ = 2.35 and [7] = 8.4 show no evidence for short periods. On the other
hand, there are three low points with log $P$ $\sim $2.6 which clearly show 
other periods around 50--60 days, allowing them alternative locations  
in the more heavily populated part of the diagram.  

There is no clear period clumping among the SRVs. Instead, there seems to be a
continuous progression, apart from the change in slope, in stellar 
luminosity, from the shortest period SRVs to the Miras. Solar neighbourhood 
SRVs with periods in the range 100-140 days show Population I kinematics 
that are similar to those of Miras with $P$ $>$ 300 days (Feast, 1963), 
although shorter-period Miras fall into population II. Also, s-process 
elements are sometimes detected in both these sets of stars (Little, 
Little-Marenin \& Bauer, 1987). This has led to suggestions that at least 
some of the SRVs are related to the long-period Miras, but pulsating in 
higher overtones.

\subsection{Amplitudes and SRV classifications}

The amplitudes of most of the SRVs are below 0.5 mag at $R$. The five Miras
with MACHO light curves have amplitudes in the range 2.5--4. A few of the 
SRVs in the range 150--200 days have amplitudes of about 1 mag. 

About two thirds of the light curves show persistent periodicity without much
change in amplitude and could be classified SRa. The remainder, although 
they usually show persistent periodicity, also show slow random or very 
long-period level-shifts and are classified as SRb. However, it should be
noted that these classsifications are subjective at best.
 
Kerschbaum \& Hron (1992) classify O-rich SRVs as ``blue" or ``red", 
according to their $V$ -- [12] and IRAS colours. Probably the blue SRVs 
correspond to those stars with [15] $\geq$ 8 and [7] -- [15] $\sim$ 0, while
the red SRVs are more luminous at 15$\mu$m and have dust emission (see
below). 

\section{Mass loss}

For an order-of-magnitude estimate of the mass loss associated with a given
star, we can estimate its 15\,$\mu$m flux excess due to dust emission by 
assuming that the photospheric flux can be extrapolated as a Rayleigh-Jeans 
tail from the 7\,$\mu$m measurement, which should be almost entirely free of 
dust emission. The result is shown in Fig.\ 5.

\begin{figure}
\centering
\includegraphics[width=0.6\textwidth]{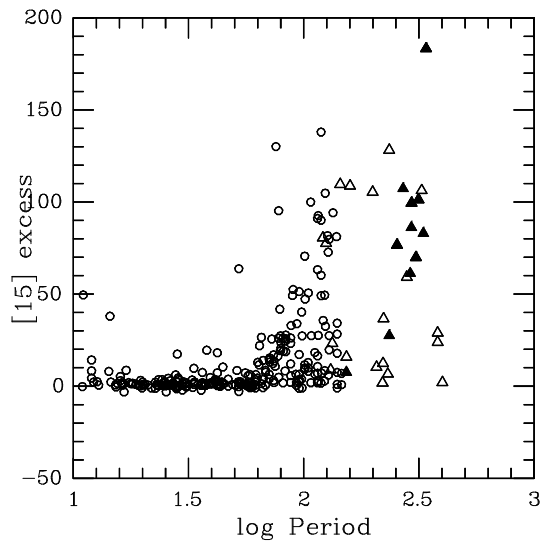}
\caption[]{Excess 15$\mu$m fluxes in mJy, beyond what is expected by 
assuming a Rayleigh-Jeans photospheric energy distribution fitted to 
the 7$\mu$m fluxes, shown plotted against log period. Having a period 
$P$ $>$ 60 days seems to be a necessary, but not a sufficient, 
condition for significant mass-loss.}
\label{fig5}
\end{figure}

The mass-loss rates for the SRVs overlap those of the shorter-period Miras
and clearly do not depend on amplitude of pulsation. The lack of measurable 
mass-loss for stars with $P$ $<$ 60 days accords with the finding of 
Kerschbaum, Oloffson \& Hron (1996) that mass-loss from stars having 
0 $<$ $P$ $<$ 75 days could not be detected in CO radio emission, while 
those in the range 75 $<$ $P$ $<$ 175 days had a 50\% detection rate.

Jura (1987) gives for the mass-loss from an AGB star:
\[
\dot{M}=1.7 \times 10^{-7} v_{15} R^2_{kpc} L^{-1/2}_4 F_{\nu,60} 
\lambda^{1/2}_{10} M_{\odot} {\rm yr}^{-1},
\]
where $v_{15}$ is the gas outflow velocity in units of 15 kms$^{-1}$,
determined from CO observations, $R$ is the distance to the star in kpc,
$L_4$ is the stellar luminosity in units of 10$^4$ $L_{\odot}$, $F_{\nu,60}$
is the flux from the object at 60\,$\mu$m and $\lambda_{10}$ is the mean 
wavelength of light emerging from the star in units of 10\,$\mu$m.
We take $v$ to be 8 km sec$^{-1}$, the average value determined for
semi-regular variables by Kerschbaum, Olofsson \& Hron (1996), $R$ $\sim$ 
8.2 kpc, and $L_4$ = 0.3, from the bolometric magnitude of a 200-day Mira 
(Glass et al.\, 1995). To relate the given 15\,$\mu$m flux to the 60\,$\mu$ flux 
required, we very tentatively take the relation by Jura (1986), intended 
for carbon stars (but see also the values of $Q_{abs}$ for astronomical
silicate grains; Draine \& Lee, 1984), namely $F_{\nu}$ $\propto$ 
$\nu^{1.54}$. If the excess 15\,$\mu$m flux is 100 mJy, we obtain 
\.{M} = 1.3 $\times$ 10$^{-7}$ $M_{\odot}$ yr$^{-1}$.

\begin{figure}
\centering
\includegraphics[width=0.6\textwidth]{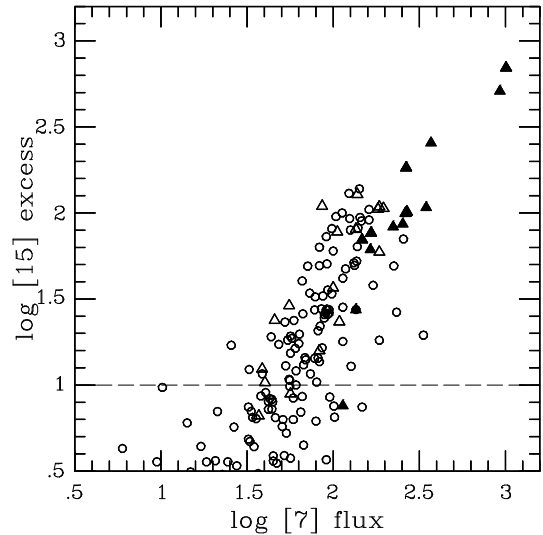}
\caption[]{Log [15] excess, an indication of mass-loss, vs log [7] flux,
an indication of bolometric mag. Below the dashed line the data may be 
subject to photometric errors, exaggerated by taking logarithms. The slope is about 2.6.}
\label{fig6}
\end{figure}

Figure 5 suggests that the lower rate limit of mass-loss that we detect 
is about two orders of magnitude less than the example discussed, or 
about 10$^{-9}$ $M_{\odot}$ yr$^{-1}$ (cf. Omont et al., 1999). 

Mass-loss in SRVs is apparently a function of $T$, $L$ and $P$. 
Because high luminosity in LPVs is also associated with low temperatures, 
it is unclear how these quantities separately affect $\dot{M}$. 
Considering $L$ as an independent variable, from Fig.\ 6 we see that 
mass-loss increases with luminosity according to the approximate relation
\[
\dot{M} \propto L^{2.6},
\]
where we have assumed that $\dot{M}$ is proportional to the 15\,$\mu$m flux
excess and the bolometric luminosity of the star is proportional to the
7\,$\mu$m flux (note that in the case of Miras dust emission may also
contribute to the 7\,$\mu$m band, leading to an over-estimate of 
photospheric luminosity).

Finally, the reader interested in the general properties of SRVs should
remember that we have discussed so far only those stars which were detected
in both MACHO and both ISOGAL bands. A preliminary glance at the light
curves of the stars seen by ISOGAL only at 7\,$\mu$m indicates that most of
them are also SRVs, but presumably with mass-loss rates too low for
15\,$\mu$m detection.

\clearpage
\flushbottom
\printindex

\end{document}